\documentclass{cflowproc}

\usepackage{natbib}

\begin{document}

\shortauthors{EILEK}     
\shorttitle{RADIO GALAXIES IN COOLING CORES} 

\title{Radio Galaxies in Cooling Cores}   

\author{Jean A. Eilek\affilmark{1}}

\affil{1}{New Mexico Tech}   


\begin{abstract}
A currently active  radio galaxy sits at the center of
almost every strong cooling core.   What effect does it
have on the cooling core?  Could its effect be strong enough to offset
the radiative cooling which should be occuring in these cores?  
In order to answer these questions we need to know how much energy the
radio jet carries to the cooling core; but we have no way to measure the
jet power directly. 
We therefore need to understand how the radio source evolves 
with time, and how it radiates, in order to use the data
to determine the jet power.  When some simple models are compared to the
data, we learn that cluster-center radio galaxies probably are energetically
important -- but not necessarily dominant -- in cooling cores.
\end{abstract}


\section{Introduction}

Cooling cores are clear observationally.  They stand out dramatically from
the general rich cluster population.  They are, however, far from
clear theoretically.  With the advent of new data, the old arguments
(where is the cooled gas?  where is the evidence for flow?) have faded,
but new ones have taken their place.

The new data are striking.  Thanks to recent, high-quality X-ray
spectra, we now know that there is simply not enough cool or cold
gas to support the old cooling-flow models.  We don't see the gas cooling
through the intermediate temperature range that we would expect from the old
models ({\it e.g.}, \cite{Fab}, \cite{Don}).  Nor do we see the
extensive reservoir of cold gas that the cooling-flow 
models predicted (although there definitely is some cold gas, 
{\it e.g.} \cite{Edge}).  Strong cooling cores are,
indeed, a bit cooler than the rest of their clusters, but only by
a factor of a few (\cite{Allen}, \cite{deGM}).  Given the short
radiative lifetime of the cooling-core gas, something must be keeping it
from cooling.

By looking at nearby cooling cores in the radio (from the VLA) as well
as in X-ray (from CHANDRA and ROSAT), we also now know that central radio
galaxies in cooling cores are interacting strongly with the plasma in the
cooling core (\cite{Blan}, \cite{Hans}).   We also know
that some cooling cores have high Faraday rotation, and thus that magnetic
fields are probably important in the cooling-core plasma (\cite{EORM}, or
\cite{Taylor};  but see also \cite{RB} for an alternative view). 
We are learning that cooling cores are complex places.

But the questions remain.  Three issues seem timely.

\begin{itemize}

\item Why are cooling core clusters different?  Is it more than happenstance
that some clusters have unusually high central densities, but otherwise
seem quite smooth and unperturbed?

\item  What controls the thermodynamics of cooling cores?  Why is the
temperature structure so regular, and what keeps the gas from cooling?

\item How important are central radio galaxies to the cooling cores?
Does every cooling core contain an important radio galaxy?  Does the jet 
deposit enough energy in the local gas to be important
in the thermodynamics of the core? 

\end{itemize}

In this paper I will focus on the last issue, the role of radio galaxies.
With an eye to the energetics of the cooling core, 
this issue can be broken up into two further questions:

\begin{itemize}

\item  What is the jet power of the central radio galaxy over the
lifetime of a typical cooling core?

\item  What fraction of that power is deposited in the gas of the cooling
core?  

\end{itemize}

In this paper I will only address the first question, which is complex
enough by itself.  I will leave the question of energy deposition to others.
The place to begin is with the data.  In \S 2 I demonstrate that almost
every strong cooling core has a currently active radio source, 
and in \S 3 I
point out that these cluster-center sources are atypical of the
broader radio galaxy population.  This is consistent with strong
interactions disturbing both the radio source and the cooling core.
But what is the jet power? Because it is not directly observable, 
we must consider how it affects what
we can observe -- the dynamics and radio power of the source. After
setting the stage, in \S 4, I explore toy models (in \S 5 and \S 6, with
important caveats in \S 7) which can connect the observables to the jet
power.  
Finally in \S 8 I discuss what we can, or cannot, definitely say about
the importance of radio jets in cooling cores.

\section{Data:  radio sources in cooling cores}

We want to know how important  cluster-center
radio sources (CCRS) are to the energy budget of a ``typical'' cooling core.
The first question to ask is how frequently the central galaxies in nearby
cooling cores have active radio sources, and what are the radio powers
of those sources.   This should be answered statistically, rather than
anecdotally, so we must consider complete samples of cooling cores.  Two
are available.

\subsection{Samples of cooling cores}

The first sample I take from the  X-ray bright (flux-limited) sample
from  \cite{Peres}.
These authors used pointed ROSAT data for the brightest X-ray clusters
in the sky.  They carried out a deprojection analysis, and from that
determined central cooling rates, cooling radii and ``$\dot M$'' mass
inflow rates.  I've taken the clusters from this set with 
$\dot M > 30 M_{\odot}$/year - these are the brightest, most centrally
peaked clusters -- and which are in the northern sky, to overlap
with the sky coverage in the  NVSS \citep{nvss}.  This gives 30 clusters.  
Comparing this sample to the existing radio data -- in the literature 
and using the NVSS --  I find that the central galaxy in 
25 of the 30 clusters has a currently active radio source, at or above
a few mJy at 1.4 GHz. 

The other sample I use is taken from the Rosat All-Sky Survey (RASS)
 of 288 nearby Abell clusters, from \cite{Ledlow}.The RASS is too photon-poor
to allow for deprojection analysis, so \cite{Ledlow} formed aperture
fluxes, quoting the 2-10 keV power within 62.5 kpc and 500 kpc of the
cluster core.  Cooling cores can be identified in this sample as those
clusters which are X-ray bright and centrally concentrated.  To form a
sample for their VLA survey, \cite{Tom} select clusters with those
properties, and which  contain a  massive central galaxy 
coincident with the X-ray peak.  Using these criteria they find a set
of 22 likely cooling-core clusters, which  has 12 clusters in
common with the \cite{Peres} sample.  Checking the literature and the
NVSS reveals, again, that most of these clusters have previously-known
central radio galaxies.  \cite{Tom} are following up on this with dedicated
VLA studies of the set.  They show that {\it all} of this sample
(22/22) have a currently active central radio source.

This is an important result:  
almost every (and probably every) central galaxy in a cooling core hosts
a currently-active CCRS.  However, this
is not necessarily due to the cooling core.  \cite{LO96} used optical and
radio data on a complete sample of
188 radio sources in Abell clusters to form  bivariate
luminosity functions.  They found that the brighter the parent galaxy,
the greater the chance of it hosting a detectable radio source.  Galaxies
in their 
optically brightest subsample have a better than half chance of having
an active AGN.   Because  the central galaxies in cooling cores
are very bright and massive -- well above $L_*$ -- it may be the
galaxy, not its X-ray atmosphere,  that causes the radio source to be
active.  

\subsection{Statistics:  relative powers}

Thus, radio sources exist in essentially every strong cooling core.  But
how powerful are they?  In order to compare radio and X-ray powers, we
must be specific about definitions. 

Many X-ray analysis papers, for instance the 
work by \cite{David} (from which \cite{Peres} take their total 
cluster powers) calculate the X-ray luminosity from the
full cluster.  This of necessity requires some assumptions about
the underlying cluster structure, such as fitting $\beta$ models to
the cluster.  On the other hand, \citet{Ledlow} take
a more conservative approach, and quote the (directly measurable)
flux within a 500 kpc aperture.  I prefer the latter and use it in this
paper;  where necessary I convert the \cite{David} values to a 500 kpc 
aperture, using clusters in common in the two samples.  Note that the
\cite{Ledlow} fluxes are within 2-10 keV;  an additional factor 
$\sim 2$ \citep{David} will convert them to bolometric. 

\begin{figure}[htb]
\plotone{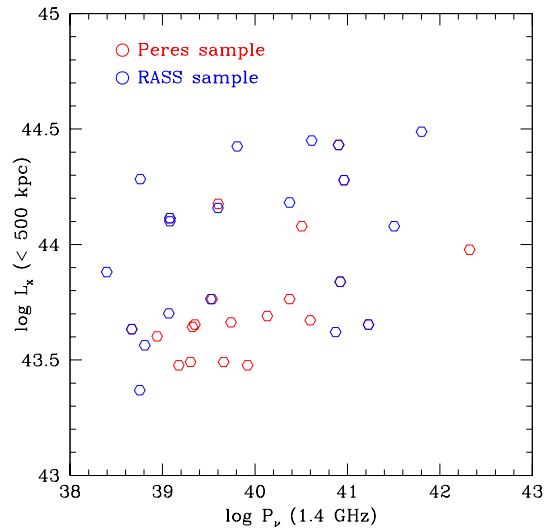}
\figcaption{Radio power at 1.4 GHz {\it vs.} X-ray power between 2 and 10
keV from the central 500 kpc of the cluster,
for both samples discussed in the text.  Units on both axes are erg/s.
 No correlation can be seen:  the
large-scale X-ray power and the radio power of the central
source do not know about each other. 
\label{Eilek:XRtot}}
\end{figure}

The radio power also merits comment.  Unlike X-ray telescopes, 
which sample a known spectrum over a broad frequency range, radio
observations only measure one frequency at a time.
Furthermore, we don't know the radio spectrum.  Synchrotron 
emission is generally a power law but also shows curvature, and 
varies source to source.  The radio spectrum is measured for some 
bright sources, but for most sources (including most of these samples)
only the flux density, $S_{\nu}$, at one frequency (1.4 GHz) is available.
I therefore  use ``power at $\nu$'', $P_{\nu} = \nu S_{\nu}$, instead of 
the ``bolometric'' $P_{rad} = \int S_{\nu} d \nu$.  

Our first quantitative result is that there is no correlation between
the X-ray power (within 500 kpc aperture) and the radio power of the
CCRS, as shown in  Figure \ref{Eilek:XRtot}.   This
isn't surprising;  the cluster is a big object and we wouldn't expect
such a correlation.  

\begin{figure}[htb]
\plotone{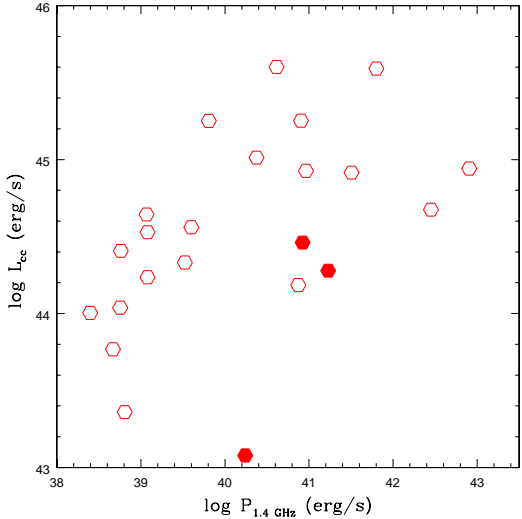}
\figcaption{Radio power at 1.4 GHz plotted against the bolometric
X-ray power within
the classical cooling radius, from the deprojection analysis of
\cite{Peres}.  This cooling radius, typically 100 - 200 kpc for
this sample, was derived for a cluster age of 13 Gyr, and is therefore
an upper limit to the likely cooling radius (as discussed in 
\cite{Peres}).
 A modest correlation is apparent between the radio power
and cooling-core X-ray power.  The solid point at the bottom is
M87, which is unusually radio-strong compared to its cooling core; the
other two solid points are 3C317 in A2052, and 3C338 in A2199.  All
three of these sources are modelled in \S 6.
\label{Eilek:Pccores}}
\end{figure}

However, because we're asking about the impact of the radio jet 
on the cooling region,  we really want to consider X-ray powers of
the cooling cores.  Using 
the data in hand, we can define the ``central region'' in two ways. 
\cite{Peres} derive cooling radii, the radius within which
the radiative cooling time is equal to the Hubble time.  The cooling radii
$\sim 100 - 200$ kpc for this sample.   The X-ray
power within this region could be called the ``cooling core'' power,
$L_{cc}$.   \cite{Peres}
point out that this is really a ``maximal'' cooling core, because the
cluster may well not have remained undisturbed for the age of the 
universe.\footnote{The relevant time scale, for instance, might better be the
time since the last major merger in the cluster -- that would give a
smaller cooling radius and smaller $L_{cc}$.}  Alternatively, 
\cite{Ledlow} give the X-ray power within 62.5 kpc, $L_{62.5}$,
 a scale which is smaller than most (maximal) cooling radii,
 but which matches the scales of the CCRS quite well.  

\begin{figure}[htb]
\plotone{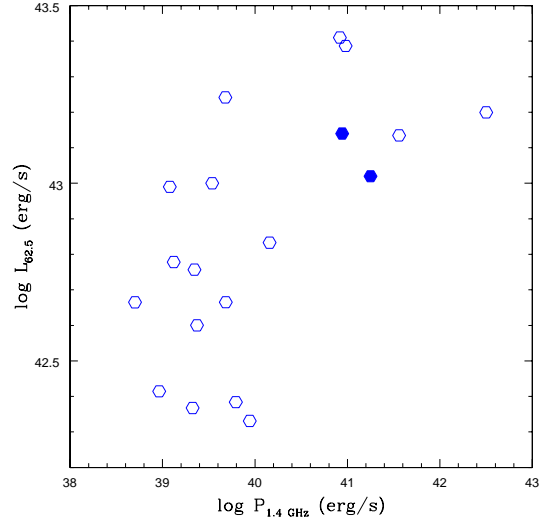}
\figcaption{Radio power at 1.4 GHz plotted against the 2-10 keV
X-ray power in the central 62.5 kpc, from \cite{Ledlow}.
  A factor $\sim 2$
converts this to bolometric power (\cite{David}). 
Once again, using this measure we find a correlation between the radio
power and the central X-ray power.  
The X-ray aperture used here is typical of, or a bit greater than, the
size of a well-developed cluster-center radio source.  The 
two solid points are 3C317 in A2052, and 3C338 in A2199. 
(M87 is not included in this sample). 
\label{Eilek:Rccores}}
\end{figure}

  Figures \ref{Eilek:Pccores} and \ref{Eilek:Rccores} compare the X-ray
core power, using both definitions, to the single-frequency radio power.
The story becomes more interesting when  we consider only the central region.  
The radio power from the CCRS does show some correlation with the core
X-ray power (\cite{Burns} noted a similar trend for a different sample). 
This correlation
may be telling us that the jet power knows about the large-scale
cooling core environment (although just how is far from clear), or it
may reflect the way in which the extended radio source reacts to the
pressure in its surroundings.

The  relative ranges of X-ray core power and radio source
power  will be useful for the modelling below. 
For the bolometric flux from the 62.5 kpc core,  the range of powers is
\begin{equation}
{\rm 62.5 ~kpc ~cores:} \quad
L_{62.5} \sim \left( 10 - 10^4 \right) P_{\nu}
\label{Eilek:ccrat}
\end{equation}
The  range of powers for the maximal cooling cores is 
\begin{equation}
{\rm maximal ~ cooling ~ cores:} \quad
L_{cc}  \sim \left( 300 - 3 \times 10^5 \right) P_{\nu}
\label{Eilek:corerat}
\end{equation}
The range of radio power is much
greater than the range of X-ray power;  this could reflect either the
range of intrinsic jet power, or the time evolution of the radio power (at
a fixed jet power), or both.

\subsection{Summary:  radio source statistics}

The central galaxy in
(almost) every strong cooling core contains an active galactic nucleus (AGN)
and a currently active, jet-driven radio galaxy.   The 
radio power of the CCRS is somewhat
correlated with the  X-ray power  the cooling cores in which they sit,
although the range of radio power is much greater than the range of X-ray
core power.  This may suggest that the strength of the cooling core has
some connection to the jet power, but also that the radio power of a
given source evolves significantly over the life of the source.

\section{Data:  CCRS are different}

Radio sources in cooling cores are different from the general population
of radio sources in galaxy clusters.  This suggests that the special
conditions in cooling cores impact the radio source therein.  In order
to understand how CCRS differ, we must review the general properties
of radio galaxies.

\begin{figure}[htb]
{\epsscale{1.1}\plotone{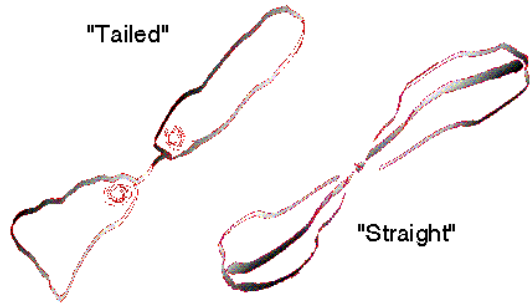}}
\figcaption{A cartoon of the two types of radio galaxy commonly found
in clusters.  Almost all of the sources in the Owen-Ledlow cluster
sample are one of these two types (or are too small to be resolved).
The evolution of each is governed by directed momentum flux from the jet,
as discussed in the text;  the differences are probably due to internal
fluid instabilities.  Radio sources in cooling cores often do {\it not} 
resemble these types, but look more like the cartoon in 
Figure \ref{Eilek:bubble}.
\label{Eilek:twotypes}}
\end{figure}

\subsection{Radio sources throughout clusters}

To understand the general nature of radio sources in the nearby
universe, we can use the work of Owen and Ledlow (1997 and 
references therein), who present radio and optical data on $\sim 250$
radio sources in  nearby Abell clusters.  Study of their images reveals
that the traditional Fanaroff-Riley Type I and Type II
classifications are not the entire answer.   Most radio
galaxies  in clusters are Type I
in terms of their radio power and optical luminosity of the parent
galaxy,  but they show a great variety of
morphologies and dynamics \citep{E02}. 

Nearly all of the resolvable Owen-Ledlow sources (about 3/4 of the
total set) can be divided into 
two morphological groups,  ``tailed'' and ``straight'', as illustrated 
by cartoon in Figure \ref{Eilek:twotypes}.  Only a few
of the  resolvable sources fit into neither category -- and they are
uniquely  located in cooling cores (as discussed below).

Straight sources are created by directed jet flows, which continue more or
less undisturbed from the galactic nucleus out to the end of the lobes.
Straight sources comprise $\sim 1/5$ of the sample, 
and typically extend 30 - 150 kpc from the galactic core.
This class includes a few classical doubles (Fanaroff-Riley Type II sources)
-- in which the jet remains very narrow and ends in a bright hot spot -- but
these are rare, only a dozen  in the entire  sample.  

Tailed sources are also created by directed jet flows, but in these
souces the jet is disturbed close to the galactic core.  It broadens
suddenly (usually), or gradually (occasionally), but
does not fully disrupt; the flow continues   on into the
characteristic tails. These tails typically can be traced $\sim 50 -
300$ kpc from the galaxy; due to surface brightness decay their
ends are often undetected. The dynamics of the tail flow are 
driven by  a mix of
momentum flux from the jet, buoyancy, and flows in the local intracluster
medium (ICM).  Tailed sources comprise $\sim 1/2$ of the Owen-Ledlow sample.

We have found no  apparent environmental reason for these two types; their
occurance is not correlated with radio power or size, local ICM density,
or magnitude of the parent galaxy. 
\cite{E02}  speculate that  the development, or not, of jet-disrupting 
instabilities is the important factor. 

It should also be emphasized  that FR Type II sources -- the famous
classical doubles -- are rare in this sample, as they are in general
in the nearby universe.  This is unfortunate; 
even though they are the only type of
radio galaxy that we understand well, they are {\it not} a good
example for studying radio sources in clusters.

\subsection{Cluster-center radio sources}

From the radio data for the two sets of cooling cores, we
learn that CCRS are not typical of
the general radio galaxy population.  They differ in three important ways.

\begin{figure}[htb]
\plotone{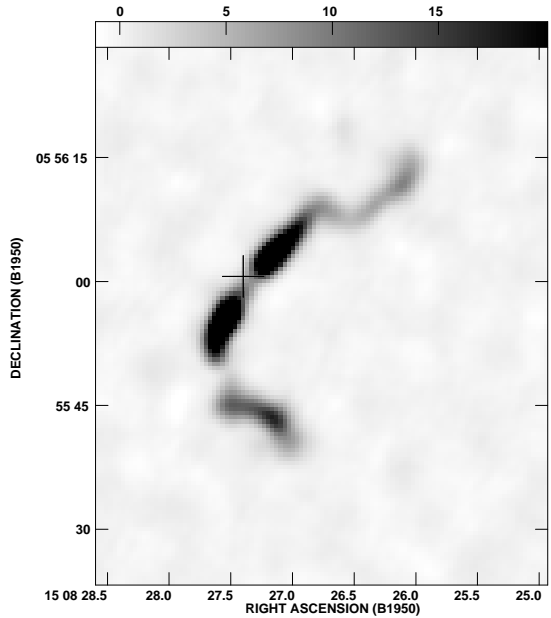}
\figcaption{The central radio source in A2029, which is
typical of the small, tailed radio galaxies found in cluster cores. 
The cross marks the position of the galactic core. 1.4 GHz VLA image from
  \cite{OL}; see also \cite{Tom} who find that the tails
are more extended and possible evidence of an underlying amorphous halo.
\label{Eilek:a2029}}
\end{figure}

\begin{figure}[htb]
\plotone{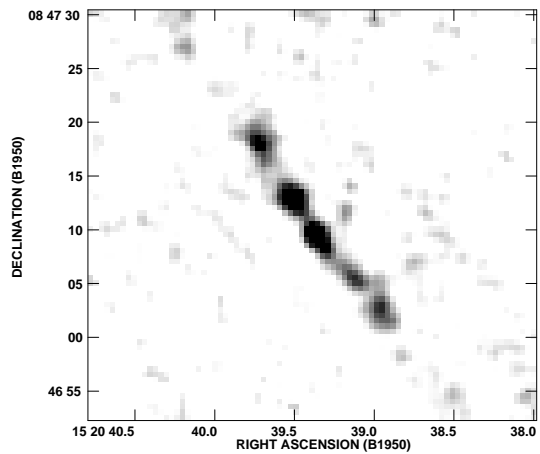}
\figcaption{The central radio source in A2063, another of
 the small, tailed sources found in cluster cores.   The source
is symmetric about the galactic core, which is in the center of the
image. 1.4 GHz VLA image from   \cite{OL}.
\label{Eilek:a2063}}
\end{figure}

\begin{figure}[htb]
{\hspace{0.5in}\epsscale{0.7}\plotone{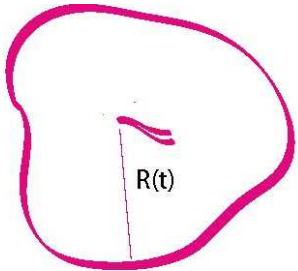}}
\figcaption{A cartoon of an amorphous, core-halo radio source;  this
morphology is common in cooling cores but rare elsewhere in the universe.
 The jet disrupts close
to the core, and injects matter and energy to the radio bubble
quasi-isotropically.  $R(t)$ is the source radius, which grows with
time; models of the source growth are discussed in \S 4.
\label{Eilek:bubble}}
\end{figure}

\begin{figure}
\plotone{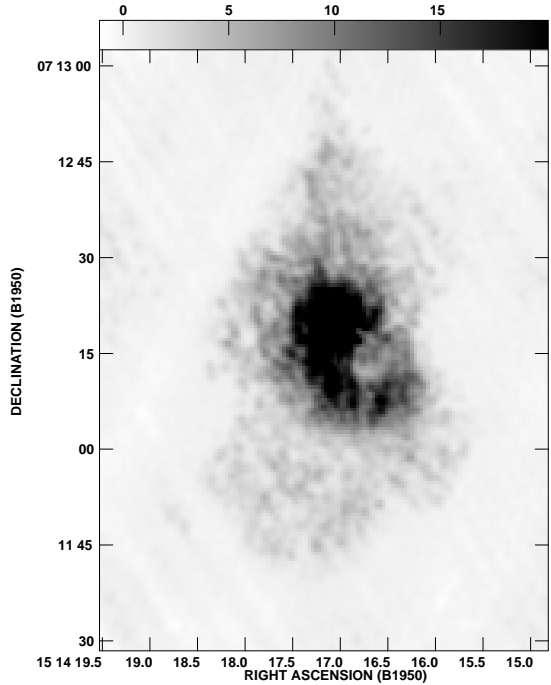}
\figcaption{3C317, the central radio source in A2052.  The center of the
galaxy coincides with the bright radio core, central in this image.  VLA
studies \citep{Zhao} show that no radio jets exist past 200 pc, although
VLB studies detect a 15 pc jet in the core. Thus, the jet is disrupted
very close to the core and feeds a quasi-spherical ``bubble'' which is
expanding into the plasma of the cluster core \citep{Blanton}. 
 1.4 GHz VLA image from  \cite{OL}; see also \cite{Tom} and \cite{Zhao}
for other radio images. 
\label{Eilek:a2052} }
\end{figure}

\begin{figure}
\plotone{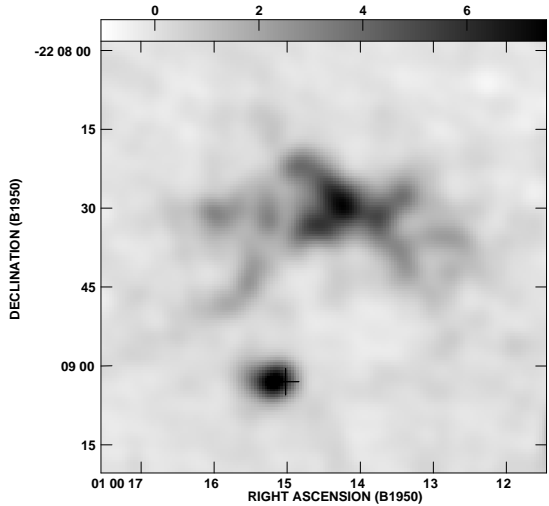}
\figcaption{The central radio source in A133, one of the ``amorphous''
CCRS.  The center of the galaxy coincides
with the bright radio core to the south of the image. 
Although this fainter source 
cannot be imaged as deeply as A2052, both the radio morphology and the X-ray
imaging \citep{Fujita}
 suggest that this is also a ``bubble''-like source, expanding
into the cooling core of the cluster. 1.4 GHz VLA image from \cite{OL};
see also \cite{Tom} for a deeper image, which establishes the connection
between the galactic core and the halo to the north.
\label{Eilek:a133}}
\end{figure}

\begin{figure}
\plotone{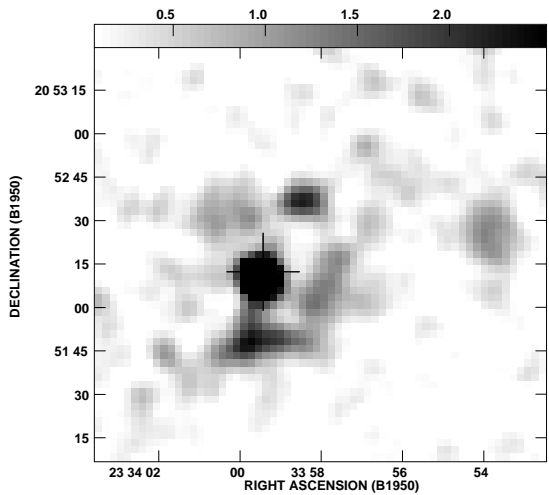}
\figcaption{The central radio source in A2626, another of the amorphous
CCRS.  The center of the galaxy coincides with the
central bright radio core;  diffuse emisison surrounds the core.  
1.4 GHz VLA image from \cite{OL}; see also \cite{Tom} for a deeper image.
\label{Eilek:a2626}}
\end{figure}

\begin{figure}[htb]
\plotone{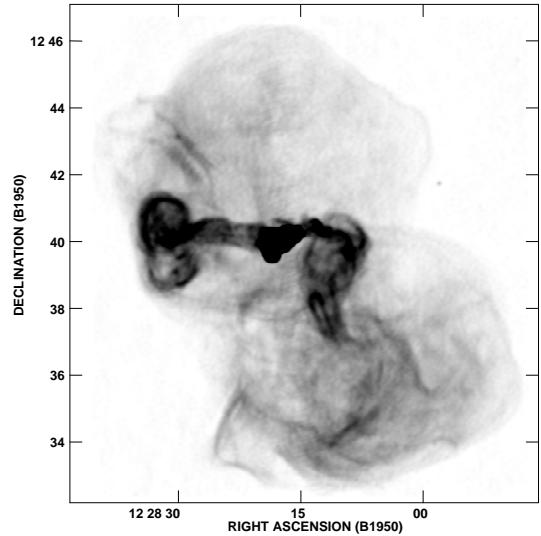}
\figcaption{The prototypical ``amorphous halo'' source M87, in the core
of the Virgo cluster.  The halo extends $\sim 35$ kpc from the galactic
core.   This high-quality image, from \cite{OEK}, makes it 
clear that the inner jet disrupts within $\sim 2$ kpc,
 but continues to feed plasma into the
radio halo. Note that the ``plumes'', which extend from the core into the
halo, are significantly brighter than the rest of the halo; if the source
were further away, and fainter, only the plumes might be detected, and
M87 would be identified as a tailed source.
\label{Eilek:M87}}
\end{figure}

\begin{figure}[htb]
{\hspace{-0.1in}\epsscale{1.2}\plotone{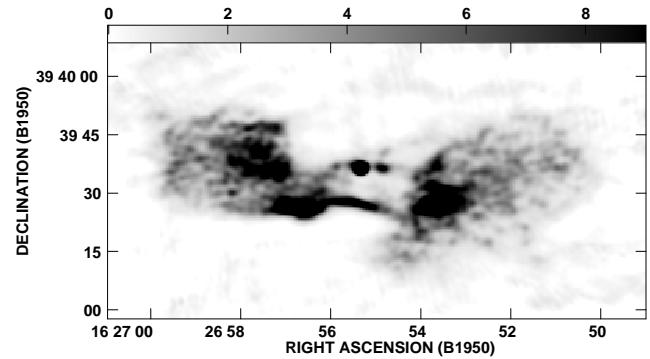}}
\figcaption{A2199, another unusual CCRS.  The radio core -- which coincides
with the galactic nucleus, and which hosts a currently active two-sided
VLB jet -- is the bright spot in the center of the
image.  The X-ray image
(\cite{OE98}) makes it clear that the radio plasma and ICM are interacting
strongly.  The filament to the south, which appears jet-like, does not
coincide with any feature in the galaxy.
It's tempting to speculate that this source has been caught in the act
of restarting, and that the southern filament is just a bright feature
arising from  ``weather'' in the radio plasma-ICM interaction.   
1.4 GHz VLA image from \cite{OL}.
\label{Eilek:a2199}}
\end{figure}

First, 
CCRS are morphologically different from the normal run of radio
galaxies.  Almost all CCRS large  enough to be resolved with current
data are either tailed sources, or unusual amorphous ones.  (Cygnus A
is the one exception, a FRII in a strong cooling core; \cite{Cyg}). Figures 
\ref{Eilek:a2029}  and \ref{Eilek:a2063} 
 show typical small tailed sources.   The
amorphous sources (as sketched in Figure \ref{Eilek:bubble}) tend 
to have diffuse haloes surrounding an active
AGN core.  Figures \ref{Eilek:a2052} through \ref{Eilek:a2199}
 show examples;  3C84 in the Perseus
cluster \citep{Blundell} is another well-known case. 
Based on current data, tailed and amorphous sources 
types occur about equally often in CCRS.  However, there are
hints that some small, tailed CCRS  may be embedded in larger, faint
haloes \citep{Tom}, so that these unusual sources may be even more
common.  (Note that M87, figure \ref{Eilek:M87}, could appear as a tailed
source if it were more distant and fainter.)
In addition, {\it all} 5 of the amorphous sources in the Owen-Ledlow
sample are in cooling cores, as are M87 and  3C84;  these are very unlikely
to be more normal sources seen in projection.   

Second,
CCRS tend to be smaller than most radio sources.  With the sole
exception of Hydra  A \citep{hyd}, the tails of which 
extend further than 300  kpc from the core, the detected extent of all other
CCRS in both samples is less than 100  kpc, and most extend less than
50 kpc from the core.  This may well be due to the higher ambient
pressure in which the CCRS find themselves, which will slow down their
spatial growth. 
 
Third, 
CCRS tend to have steeper radio spectra than most radio sources (\cite{Ball},
 also \cite{Tom}).   Steep-spectrum radio emission is usually
thought to mean the relativistic electrons have suffered significant
synchrotron losses; but other interpretations, such as strongly
inhomogeneous magetic fields, are also possible.

\subsection{Summary:  cluster-core radio sources}

The unusual nature of CCRS strongly suggests that something special in
the cooling-core environment affects the growth and nature of the
central radio source.  The small size, and unusual morphology, hint
that the cooling-core environment slows the growth of the source, and
may fatally disrupt the jet very close to the AGN.  The isotropized
energy flow from the jet would create an amorphous halo, rather than
a tailed radio source.  The jet disruption by the local cooling-core
plasma would
also be expected to disturb and heat the plasma.  This picture
is, of course, consistent with the evidence from CHANDRA of strong
interactions between the radio source and the plasma in some
cooling cores.

\section{Interlude:  the jet power}

We  now know that radio sources are common, and probably
universal, in the bright galaxies which sit centrally in cooling cores.  
We have strong evidence that CCRS are unusual, and are interacting 
strongly with the local ICM.  
This interaction must
perturb the local ICM as well as disturb the radio galaxy;  some
amount of the jet energy must be deposited in the local ICM.  What is
not yet clear, however, is how important this interaction is to the
energy budget of a ``typical'' cooling core.

In order to address this question we must determine the strength of
the jet in a ``typical'' cooling-core AGN.  The total energy flux in a
radio jet of speed $\beta c$ and cross section $S = \pi r^2$ is
\begin{equation}
P_j = \pi r^2  \gamma^2 \beta c \sum_{i,e} \left( \rho c^2 + 4 p \right)
+ { c \over 4 \pi} \int_S d \bold s \cdot \bold E \times \bold B 
\label{Eilek:Pjet}
\end{equation}
In most usage, $\bold E$ is taken as the inductive field: $\bold E = -
\bold v \times \bold B / c$. 
For later use, we can rewrite equation \ref{Eilek:Pjet}
as 
\begin{equation}
P_j = P_{je} + P_{ji} + P_{jB}
\end{equation}
which generically represents the flux  of relativistic electrons, ions, and
magnetic field.   We will see below, that the radio power 
measures $P_{je}$;  the mean magnetic field
 in the source at time $t$ can be related
to the ratio $P_{jB} / P_j$;  and we have no {\it a priori} knowledge of the
ion flux, $P_{ji}$.

The best way to determine $P_j$ is of course to measure it directly.
At present this is possible for only one source -- M87 -- and only as
a lower limit in that case \citep{OEK}. This jet is resolved, so we know
its radius;  proper motion studies find $\gamma \sim$ a few for bright 
features, and physical analysis tells us the flow speed is unlikely
to be much slower than the feature speed \citep{Phil}.  Finally, from
standard synchrotron analysis, we know the {\it minimum} pressure in
the jet;  this is the lowest value of $p_B + p_e$ consistent with the observed
radio emission.\footnote{It is important to remember that
``equipartion'' (and minimum pressure) are really measures of the
synchrotron emissivity, which $\propto p_B p_e$ -- although they are
often interpreted as measurements of the energy (or pressure) in
electrons and field separately.} For the M87 jet, these methods find
$P_j \gtrsim {\rm few} \times  
10^{44}$erg/s.  For comparison, the X-ray luminosity of the maximal
cooling core $L_{cc} \sim 3 \times 10^{43}$erg/s \citep{Peres}.  Thus,
this jet is clearly energetically important to the core of this
cluster.  

For other sources, we must use more indirect -- and more model-dependent 
-- methods.  I discuss two in this paper.  One method is to connect the
radio power to the jet power.  Because each source varies with time,
this cannot be done for an individual source, but will have some statistical
validity.  Another method is to construct dynamical models of the source
evolution, and from them connect the source size and internal energy to
the jet power.  Both of these methods require assumptions, in order to
build the models;  they must be checked, {\it post hoc}, for the
validity of these assumptions.  The two methods are also interconnected.
Modelling the radio power requires knowledge of the source dynamics.
Extracting the jet power from the source dynamics requires independent
evidence of the age, which comes in principle from the radio spectrum. 

In the rest of
this paper I highlight the assumptions and general results of the models,
focusing on what they can tell us about heating of cooling cores. 
Both models will be described in more detail in forthcoming papers
\citep{E04}.

\section{Toy models:  evolution of radio power}

One first thinks of using the radio power to measure the jet power.
We know the synchrotron power depends on relativistic electrons and magnetic
field in the source.   As the source evolves, so
should the radio spectrum.  Although the detailed plasma physics can be
very complicated, it seems best to start with a simple model, and see
what it predicts.  

I therefore take the traditional approach to the relativistic 
electrons, namely,  assuming that the jet injects new particles
 at a steady rate $q(\gamma)$ [normalized so that 
$P_{je} = \int q(\gamma) \gamma m c^2 d \gamma $].
If the electrons feel no further acceleration in the lobe, but simply lose
energy at a rate $d \gamma /d t$, the electron distribution evolves as
\begin{equation}
{ \partial n (\gamma) \over \partial t} + { \partial \over \partial \gamma}
\left[ n(\gamma) { d \gamma \over dt} \right] = q(\gamma)
\label{Eilek:DF}
\end{equation}
I specify to synchrotron losses, $ d \gamma / dt \propto \gamma^2 B(t)^2$.  
(Adiabatic losses are also important for young sources and low-$\gamma$ 
particles, but synchrotron losses will soon dominate).  Looking ahead to
the dynamical models, I assume a magnetic field which slowly decays with
time,  $B(t) \propto t^{-1/6}$ (different decay rates change
the details of the arguments below, but not the substance).   Again
following tradition, I assume that the jet  injects a power law in the
energy range $\gamma_o < 
\gamma < \gamma_m$.  Solutions of equation \ref{Eilek:DF} are
straightforward \citep{ES}, giving a broken power-law
electron spectrum which steepens at a critical energy $\gamma_c(t)
\propto 1 / B^2 t  \propto t^{-2/3}$.  The low-$\gamma$ distribution  
 mimics the injection spectrum, and the 
high-$\gamma$ distribution steepens due to synchrotron losses.  

\subsection{Solution: radio power of one source}

To determine the radio spectrum, note that the electron
energy which ``maps'' to the observing frequency $\nu$ is 
\begin{equation}
\gamma(\nu)
\propto [ \nu / B(t) ]^{1/2}
\label{Eilek:synu}
\end{equation}
 and the  electron distribution maps to the synchrotron spectrum as 
\begin{equation}
S(\nu,t) \propto B(t) \gamma(\nu)^{1/2} n[\gamma(\nu),t]
\label{Eilek:synch}
\end{equation}
Applying equation \ref{Eilek:synch} to the solutions of \ref{Eilek:DF},
and assuming that the magnetic 
field doesn't fluctuate too much in space -- taking $B(t)$ as uniform
across the source -- the synchrotron spectrum is also a broken power
law, steepening at the critical frequency, 
\begin{equation}
\nu_c(t) \propto \gamma_c(t)^2 B(t) \propto t^{-3/2}
\label{Eilek:crit}
\end{equation}
To be specific, I take  $q(\gamma) \propto \gamma^{-2}$.  The solution
to \ref{Eilek:synch} then has two parts.  We would call a source
``young'' if it is observed when  $\nu_c(t)$ is above our
observing frequency (early times, flatter spectrum, low frequencies).  
For these sources, the synchrotron flux obeys
\begin{equation}
S_{\nu} \propto t^{3/4} \nu^{-1/2} ~; \quad \nu < \nu_c(t)
\end{equation}
Alternatively, a source observed when $\nu_c(t)$ is below our
observing frequency would be ``old'' (later times, higher frequencies, 
steeper spectrum).  For these sources, the synchrotron flux obeys
\begin{equation}
S_{\nu} \propto t^{-3/8} \nu^{-5/4} ~; \quad \nu > \nu_c(t)
\end{equation}
The high-$\nu$ spectrum is steeper here than in the standard
solutions, because the $B$ field decays with time.  

\begin{figure}[htb]
\plotone{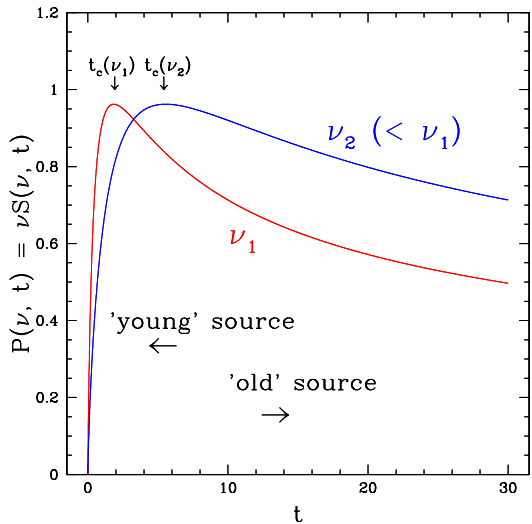}
\figcaption{The time evolution of a radio source, seen at two different
frequencies.  At early times (when the source is ``young''),
the source brightens, as the jet supplies more
electrons and field to the source.  When the age of the source reaches the
synchrotron lifetime, and becomes ``old'' the source
begins to fade, as radiative losses overcome the ongoing supply of new 
electrons.  Because the synchrotron life is a function of particle energy,
and thus observing frequency, a source can be young at a low frequency and
old at a high frequency. 
\label{Eilek:timepower}}
\end{figure}

\begin{figure}[htb]
\plotone{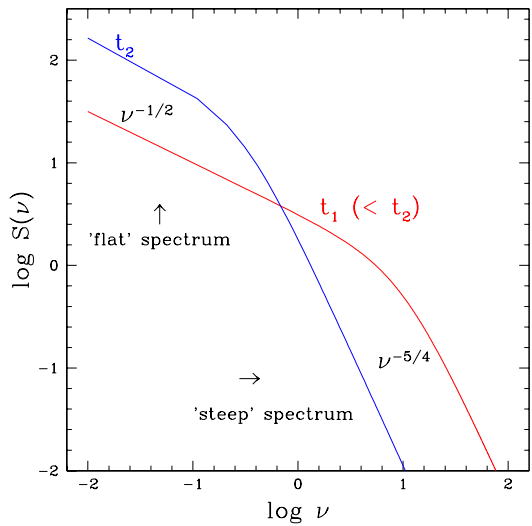}
\figcaption{The spectrum of radio source, seen at two different times.  
At low frequencies the source appears ``young'', with a flatter spectrum;
at high frequencies the source appears ``old'', with a steeper spectrum.
Note the steeper high-frequency spectrum here, compared to standard
spectral-aging models -- due to the decay of the mean magnetic field
with time as the source expands.
\label{Eilek:timespec}}
\end{figure}

Thus, a young
source brightens with time (as more and more electrons fill the
source).  But an old source  decays slowly with time (as synchrotron
losses offset the 
ongoing input of new electrons).  
A source reaches its peak power when $\nu_c(t)$ equals one's observing 
frequency.  This provides a useful definition of the synchrotron lifetime
of the source:  $\nu_c(t_c) = \nu$.  Illustrative solutions are shown
 in Figure  \ref{Eilek:timepower} and in Figure \ref{Eilek:timespec}. 
The peak single-frequency power the source attains is
\begin{equation}
\max (P_{\nu} )\simeq { 0.5 P_{je} \over \ln (\gamma_m/\gamma_o)} \sim .05 P_{je}
\label{Eilek:maxP}
\end{equation}
In the last I have relied on  typical injection models which might 
have $\gamma_m \simeq (10^4 - 10^5 ) \gamma_o$.  Although the local $B(t)$
 field is important to the amplitude of $P_{\nu}$ away from its maximum,
the peak $P_{\nu} = \nu_c S(\nu_c)$ depends only on $P_{je}$.
Thus the peak radio power is a good measure of the electron power in
the jet.

\subsection{Statistics: how the population evolves}

\begin{figure}[htb]
{\epsscale{1.2}\plotone{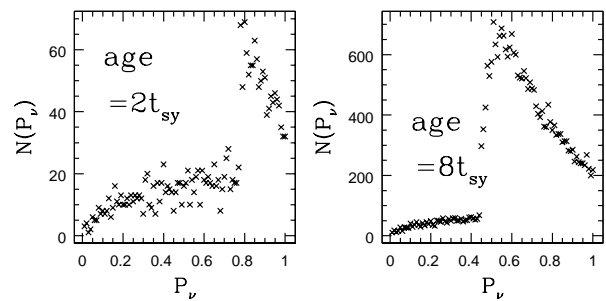}}
\vspace{-1.8in}
\figcaption{The time evolution of a population of cluster-center radio
galaxies.  In this simple model,  sources are created at a steady rate, 
each starting life at zero radio power;  all sources 
have the same intrinsic jet power and the same synchrotron lifetime.  The
observed variation is entirely due to evolution of individual sources.
The horizontal axis is normalized to $\max{P_{\nu}}$, from equation 
\ref{Eilek:maxP}.
\label{Eilek:LFs}}
\end{figure}

We do not, however, follow one source with time;  
rather, we sample a population containing a range of ages.  Once $P_{\nu}(t)$
is known, this can easily be accounted for.  If
sources are created with zero power,  the population obeys
\begin{equation}
{ \partial N(P_{\nu}) \over \partial t} + { \partial \over \partial P_{\nu}}
\left[ N (P_{\nu}) { d P_{\nu} \over dt} \right] = 0
\label{Eilek:popn}
\end{equation}
In a steady system, this gives $N(P_{\nu}) \propto 1 / ( d P_{\nu} / dt)$.  
This has two branches, for young sources which brighten with time,
and for older sources which grow fainter with time;  a
simple illustration of this is in Figure \ref{Eilek:LFs}.
Thus, if the toy model used here -- steady creation
of new sources and constant jet power in those already born --
continues to operate, 
 the population becomes dominated by older, steep-spectrum sources
which decay only slowly with time.   The steep-spectrum sources,
being older, will be larger than the smaller (younger) flat-spectum sources. 

How does this toy model compare to the CCRS samples we have?  Again looking
ahead to \S 6, where I find synchrotron ages to be much less than the 
age of the cluster, we would expect almost all of the CCRS samples to 
contain large and steep spectrum sources.  This is not the case;  while
many CCRS do match this description, many are still smaller and flat 
spectrum.  Most of the CCRS population seem to be no more than a few
 $ t_{c}$ old.  (I return in \S 7 to what this requires of the duty cycle
of the central AGN).  Referring to Figure \ref{Eilek:LFs}, 
the mean or ``typical'' radio power of a CCRS will be somewhat less
than the maximum possible: 
\begin{equation}
\langle P_{\nu} \rangle \sim 
\left( {\textstyle{ 1 \over 2} - {2 \over 3}} \right) \max( P_{\nu})
\sim O(10^{-2}) P_{je}
\label{Eilek:meanpnu}
\end{equation}
That is, the mean $P_{\nu}$ of a sample of sources will be a few per
cent  of the electron power in the jet.

\subsection{Summary:  radio power}

The most important point here is qualitative:  a source will
evolve in radio power even while its jet power stays constant.   Young
sources will grow brighter with time, and old sources will slowly fade.
  It follows that the radio power is an imperfect tracer of
the jet power.  For a particular source we must know much more than the
current-epoch radio power if we want to learn about the jet.

We can, however, make statistical statements.  From the known evolution 
of the $P_{\nu} / P_{je}$ with time, we can use the distribution of
radio powers of a sample of sources to learn about  $P_{je}$. 
I find that $P_{je} \ll L_{62.5}, L_{cc}$ for nearly all CCRS
in the two
samples in hand. Thus, the power in the electron component of the 
jet is insignificant compared to the X-ray cooling rate in nearly all cases.

However, the jet contains more than electrons.  We know it transports
magnetic energy, and it may contain other particle species as well.
Can anything be said about the other components in these jets? 

\section{Toy models:  Dynamics}

We can measure more than the single-frequency radio power of the
source.  With good radio images we can learn its morphology and  
its size. We can measure its radio spectrum;  for the brightest sources
the spectrum has been measured over several decades in frequency,
which allows comparison to the predicted spectral steepening (as in Figure
\ref{Eilek:timespec}).  The magnetic field cannot be measured directly, but 
decent indirect estimates  can be obtained by combining 
radio and X-ray data (minimum pressure, ambient X-ray pressure,
Faraday rotation).    
With this information in hand, we can use dynamical models to explore
the total jet power. 

\subsection{Radio sources as bubbles}

Once again, the place to start is with the simplest plausible model.  For
amorphous CCRS, this is indeed simple. We can describe the radio halo as a
quasi-spherical bubble, fed by a jet whose power $P_j$ stays constant
in time. If the jet disrupts close to the galactic
core, but continues to be powered by the AGN,
it will deposit its mass and energy more or less isotropically into 
the surrounding plasma.  This plasma will respond by expanding into
the plasma of the cluster core; if the expansion is subsonic relative to
the plasma -- as is likely for these sources -- the internal and external
pressures will approximately balance.  
As long as the energy input $P_j$ stays
constant, the expansion rate of the bubble will be determined by mean
energy density within the bubble, and the pressure structure
of the local cooling core.  (More details are given in
\cite{OE98} and \cite{OEK}).   Figure \ref{Eilek:bubble} shows a
cartoon illustrating 
this.   The radius of the bubble is governed by
\begin{equation}
4 \pi R^2 p(R) { dR \over dt} = { \Gamma -1 \over \Gamma} P_j
\label{Eilek:bubbleR}
\end{equation}
if the plasma in the bubble has an adiabatic index $\Gamma$. 
If the external pressure falls off as $p(R) \propto R^{-3/4}$ (typical
of cooling cores), the bubble grows as 
\begin{equation}
R(t) \propto \left( { P_j t \over p_o } \right)^{4/9}
\label{Eilek:bubbleR1}
\end{equation}
where  $p_o$ is the ICM pressure at some fiducial radius $r_o$.

The internal energy, $E_i$, in species $i$ is determined by
\begin{equation}
{ d E_i \over dt} = P_{ji}  - p_i { d V \over dt} - \Lambda_i
\label{Eilek:intE}
\end{equation}
for species $i$, which can have radiative losses $\Lambda_i$, and
 which can be plasma (with $E_i = p_i V / (\Gamma_i -1)$; 
I follow ions and electrons
separately), or magnetic field (with $E_B \simeq p_B V$ for a tangled 
field).  Once we have the $R(t)$ solution, from equation \ref{Eilek:bubbleR1},
we know $V(t)$ and equation \ref{Eilek:intE} gives us the B field directly:
\begin{equation}
{ B^2(t) \over 8 \pi } 
\propto { P_{jB} \over P_j} \left( P_j t \right)^{-1/6}
\label{Eilek:Bt}
\end{equation}
This shows that  $B(t)$ {\it drops} with time, due to the density ramp into
which the source expands, which causes $V(t)$ to grow  faster than
linearly with time.  

It should be noted here that this model describes only 
the {\it mean} field, throughout the bubble;  it does not describe
the inhomogeneous magnetic structure (due to filaments, internal flows,
and so on) which we know must exist.   This model also does not consider
the more complex physics which will become important later in the source's
evolution.  It seems likely that instabilities -- Rayleigh-Taylor, Parker,
magnetic tearing and reconnection -- will eventually mix the radio plasma
with the ambient ICM.  In addition, once instabilities (or simply
asymmetry in the local ICM) break the simple spherical symmetry, bouyancy
will also affect the structure of the source.  Comparing the radio and
X-ray data, it looks as though some sources ({\it e.g.} 3C84 in Perseus,
\cite{Blundell}, or 3C317 in A2052, \cite{Blanton})
 have not yet mixed significantly with the local ICM,
but that others ({\it e.g.} M87, \cite{NB}, \cite{Young}) are well mixed.

\subsection{Apply to real sources}

This simple model can be 
tested against two classic bubble sources, 3C317 (figure \ref{Eilek:a2052}) 
and M87 (figure \ref{Eilek:M87}),  and also for the unusual source
3C338 (figure \ref{Eilek:a2199}), which is probably also a halo-type source,
distorted by local ``weather''.  For each of these we 
 have good enough radio images to locate
the outer edge of the source, and good X-ray information exists on the 
pressure distribution of the ICM in the cluster core.  We also have
(indirect) evidence on the amount of mixing of radio and cooling-core plasmas;
based on the detection, or not, of X-ray ``holes'', we can speculate that
the plasmas are well mixed in M87, but not in 3C317 or 3C338. 
Finally we  have radio-derived minimum pressures and Faraday rotation 
information. 

From these data we can
 use equation \ref{Eilek:bubbleR1} to find the product $P_j t$;  
and compare the mean field from \ref{Eilek:Bt} to the field
necessary for pressure balance with the ICM, to estimate $P_{jB}/ P_j$.
Scaling to $10^{44}$ erg/s, 
I find $P_{44} t_{Myr} \sim 100-700$ for the three sources.  
It seems likely that the magnetic field provides a
substantial part of the pressure support of the bubble; if this is
the case,  the ratio $P_{jB} / P_j \sim 1/10 - 1/2$ in all three sources.  

\subsection{Radio spectra and source ages}

How can we go further?  In order to break the
 $(P_j,t)$ degeneracy, we need independent information on the age.  
If we accept the traditional injection-plus-aging model of the radio
spectrum, developed above, we can use the critical frequency (equation
\ref{Eilek:crit}) to get new information.  Collecting the results from
the spectral and dynamic analysis, this model says
\begin{equation}
\nu_c(t) \propto  { P_j^{1/2} \over t^{3/2} }
\left( { P_j \over P_{jB}} \right)^{3/2}
\label{Eilek:nucrit}
\end{equation}
This addition completes the set.   We can use it together with the results
above to determine $P_j$ and the source age independently.

In M87, this analysis gives  
$P_j \sim 6 \times 10^{44}$erg/s, which is a 
bit above the minimum-$P_j$ estimate from \cite{OEK},
and well above the cooling-core power in this (X-ray weak)
cluster. For 3C317 and 3C338, this analysis finds jet 
powers an order of magnitude higher than for M87. These are stronger cooling
cores than Virgo, and the factor $\sim 10$ higher ICM pressure is what
leads to the higher $P_j$ here.  If this model is right, these sources
also have jet powers  well above the X-ray luminosity of their cooling
cores.

\subsection{Summary: dynamics} 

The unusual, quasi-spherical core-halo structure of many CCRS can be
easily modelled as a ``bubble'', driven by energy input from the jet, and
 expanding against the pressure of cooling core.  From the size of such
sources we can derive a robust estimate of the product $P_j t$. From
knowledge of the magnetic field inside the bubble we can estimate the
fraction of the jet power that is carried electrodynamically:  $P_{jB} 
/ P_j \sim 0.1 - 0.5$ seems to be the case. 

To go further, and learn the jet power, we need a separate estimate of 
the source age.  Using the traditional approach, spectral aging, I find
that the three sources considered here must be quite young (a few Myr), 
in order to keep synchrotron losses from being too important.   Combining 
this with the dynamical estimate of $P_j t$ requires powerful jets.
Because this spectral aging model may not fully describe the physics of
these sources -- as I discuss below  -- the jet power derived this
way is probably an upper limit, say a ``maximal'' jet power.
Comparing to the results in \S 5, 
I find $P_j \gg P_{je}$;  the electrons are only a small component
of the jet power.  This analysis also suggests that the total jet 
power can be significant compared to the X-ray power of the core.

\section{Toy models: critique} 

These conclusions are attractive --- but they
are only as good as the theories in which they are based.  Are these
models good enough?  Three caveats are in order regarding the toy
models I have presented here.

\subsection{Not all sources are simple bubbles}

Not all CCRS can be described as homogeneous bubbles.  One complication is
that some sources (such as Hydra A and
Cygnus A) clearly retain the identity of their jet.  Others are
complex;  they appear tailed, but may also contain a large, faint
halo.  One example is the CCRS in A2029 (Figure \ref{Eilek:a2029});
another is M87 itself (Figure  \ref{Eilek:M87}), which might be called a 
tailed source if it were fainter and seen in a less deep image.  

Tailed sources
 can be modelled, using the methods of \S 5, but necessarily with
more complexity. From observations of similar, but brighter, sources in
the general population, we know that the  jet retains at least partial 
coherence while  propagating through cluster gas.  It distorts close to 
the core, forming a ``tail'',  but the flow continues on into the tail. 
The length of the tail is determined mainly by
its momentum flux (as the thrust from the directed flow pushes out against
the ICM pressure), but buoyancy and flows in the ICM may also be important.
 In general the tail length  grows less rapidly than the flow speed 
within the tail, so that
plasma reaching the end of the tail must slow down, move aside, and
be stored in a larger region (the ``cocoon'' of the tail flow).  
Growth of this cocoon will be governed by energy and mass conservation.

The qualitative results of these tailed-source
 models are similar to the bubble models
above (although different in detail).  The volume of the tail will
grow faster than linearly with time, so that the mean magnetic 
field will decay with time.  Thus the source will initially brighten in
the radio, and later slowly fade, as do the simpler bubble models. 
This issue, while very interesting from the point of view of radio
source physics, should not affect the general arguments here about CCRS
jet power and evolution.

\subsection{Is spectral aging right?}

A more serious problem with the simple models in this paper is their
reliance on standard models of the relativistic electron evolution.
The weakest links in the models of \S 5 and \S 6,
 in my opinion, are the assumptions that (i) relativistic electrons
do not undergo {\it in situ} acceleration after they are ``injected'' by
the jet, and (ii) that they radiate 
 in a magnetic field which is uniform throughout the source.  This
set of assumptions  predicts,
as we saw in \S 5, rapid decay of the high-energy electrons and rapid
steepening of the radio spectrum.   Requiring the frequency at which the
spectrum steepens to be as high as it's observed to be, in the three
sources under consideration, forced those sources to be quite young.
This in turn required very high jet powers, in order for the sources
to grow to their present size against the high ambient pressure of the
cooling core.

Detailed studies of other radio galaxies have pointed out that this
standard, simple picture is not necessarily right.  The high frequency
steepening commonly observed may be due to quite different physics
\citep{LarryR}.  What else can explain spectral steepening without
requiring such a young source?  One possiblity is {\it in situ}
particle acceleration (say by shocks or plasma turbulence) which keeps
the electrons energized despite ongoing synchrotron losses.  Another
is the effect of magnetic fluctuations on the radio spectrum;  even
though the mean field may be uniform throughout the source, small-scale
magnetic fluctuations can have a strong effect on the synchrotron spectrum
(due to the $\gamma^2 B$ dependence of the emitted frequency, equation
\ref{Eilek:synu}; {\it c.f.} \cite{EA} for specific models including
broken power laws).

\subsection{What about duty cycles?}

It seems unlikely that the radio jet in a CCRS remains at constant
power throughout the life of the cluster.  Dynamical models of specific
sources suggest they are only several tens of Myr old.  
Furthermore, not all of
the CCRS have the steep radio spectrum which the simple models predict;
this is consistent with existing sources being only a few times their
synchrotron age.  Both of these arguments point to the AGN switching into
a low-power state after something like 10-100 Myr at high power.  However,
because nearly every bright galaxy in a cooling core hosts a currently
active AGN, the low-power state cannot last for long.  The low-power
cycle must last long enough for the large ``relic'' radio source 
to fade,\footnote{Just how that can happen quickly is not clear;  one 
possibility is turbulent dissipation of magnetic field.} 
but not as long as it
spent in its high-power state.  When it again becomes a high-power
jet, the cycle would restart, and a new ``young'' radio source would
begin to grow.

It must be emphasized that the above argument is very uncertain.  We
do not know how relic radio sources behave -- in fact their relative
scarcity in the universe is a major problem for the radio source
models in general, including those
presented here.  And we have no idea what controls the duty cycle
of the AGN itself.   However, for the purposes of this paper we can
note that the estimates of typical jet power in a sample of CCRS should
probably reduced by some factor, not too large, compared to those
derived in \S 4 and \S 5 (which were based on constant jet power).

\section{Concluding Remarks}

My focus in this paper has been the nature of the radio galaxies which sit in
strong cooling cores, and their role in the energetics of the cooling core.
To that end, I reviewed the data and also discussed the physics of radio
source evolution.  In order to know how important these radio sources
are to the cluster core, we need to know the power carried by their jets.
Because we cannot measure that power
  directly, we need to understand the physics 
of the radio sources in order to use the data to estimate the jet power.  
Two important conclusions emerge.

First, we have learned that central AGN with associated 
radio sources are common, perhaps universal, in strong cooling cores.
Many of these central radio galaxies are interacting strongly
with their surroundings, and must be energizing their surroundings to some
extent.  The central AGN seem
to undergo high and low power periods, with a cycle time $\sim 100$ Myr,
but with most of the duty cycle spent in the ``high'' phase.  It is
important to realize that the radio power we measure is not a good measure
of the jet power, for any given source, because the radio power varies
significantly over the lifetime of the source even if the jet power remains
constant.

Second, we have seen that models of radio source evolution can be used to
constrain, but not uniquely measure, the jet power, $P_j$.  The mean radio
power in the sample provides an estimate of that part of $P_j$
carried by relativistic electrons, $P_{je}$.   For nearly all of the CCRS
this power is small compared to the X-ray power of the cooling core.
 This is probably a lower limit to the true jet
power.  It is interesting to note, however, that
$100 P_{je} \gtrsim L_{62.5}$ for all of the RASS sample, and
$100 P_{je} \gtrsim L_{cc}$ for 2/5 of the Peres sample.  If particle
acceleration in these AGN behaves similarly to cosmic ray acceleration in
our galaxy, there could be substantially more energy in ions than electrons
(up to the factor $\sim 100$ for galactic cosmic rays), making the
jets energetically important to many cooling cores. 

Turning to constraints on the total jet power (in electrons, 
ions and magnetic field), dynamical models
of the sources give us an estimate of the product $P_j t$.  If we also 
assume the observed radio spectral breaks are due to simple synchrotron
aging, we gain an independent estimate of the source age, and thus of
the jet power.  When applied to three well-studied CCRS, this analysis
suggests the total jet power can 
be quite large, $P_j \gtrsim L_{cc}$.  Because the spectral aging argument
is not strong, this estimate is probably an upper limit to the true jet power;
but it does suggest that the jet contains more than just its electron 
component. 

In summary, it seems likely that central radio galaxies play an important
role in the energetics of at least the inner region of the cooling core.
However, because current models cannot really pin down the jet power, the
preceding statement can be only qualitative.  In addition, because the
density and temperature profiles are so uniform in all strong cooling cores,
it seems unlikely that these short-lived, rapidly evolving central radio
galaxies control all of the  physics in these cores.  But they are
probably part of the answer to the questions posed in the introduction.


\acknowledgements
Extensive discussions with Frazer Owen and Tomislav Markovi\'c have
been invaluable in stimulating and focusing this work.  I also thank them,
and Mike Ledlow, for generously giving me full access to their data (some
well before publication).  Insightful questions from Robert Laing and
Larry Rudnick have  helped along the way.  Some of this work was done
during my sabbatical visits to the University of Oxford, and the Instituto
di Radioastronomia in Bologna;  I thank both institutions and the 
people in them for their
support.



\end{document}